\documentstyle[12pt]{article}
\input epsf
\begin{document}

\newcommand{\eq}{\begin{equation}}
\newcommand{\en}{\end{equation}}
\newcommand{\gsi}{\,\raisebox{-0.13cm}{$\stackrel{\textstyle>}
{\textstyle\sim}$}\,}
\newcommand{\lsi}{\,\raisebox{-0.13cm}{$\stackrel{\textstyle<}
{\textstyle\sim}$}\,}
\newcommand{\etagluino}{\eta_{\tilde{g}}}
\newcommand{\gluino}{\tilde{g}}
\newcommand{\photino}{\tilde{\gamma}}
\newcommand{\bino}{\tilde{b}}
\newcommand{\tsquark}{\tilde{t}}
\newcommand{\wino}{\tilde{w}}
\newcommand{\mtilde}{\tilde{m}}
\newcommand{\higgsino}{\tilde{h}}

\rightline{RU-94-70}
\rightline{hep-ph/9408379}
\baselineskip=18pt
\vskip 0.5in
\begin{center}
{\bf \LARGE Light Gluinos\footnote{Invited talk at Quarks-94,
Vladimir, Russia, May 1994.}\\ }
\vspace*{0.7in}
{\large Glennys R. Farrar}\footnote{Research supported in part by
NSF-PHY-91-21039} \\
\vspace{.05in}
{\it Department of Physics and Astronomy \\ Rutgers University,
Piscataway, NJ 08855, USA}

\end{center}
\vskip  0.5in

{\bf Abstract:}
Gluino and lightest neutralino masses are naturally less than
a few GeV if dimension-3 susy-breaking operators are absent from the
low energy theory.  In this case gaugino masses come from loops
and are calculable in terms of known particle masses and two mass
parameters, $\mu$ and $\tilde{m}$.  The phenomenology of such a
scenario is discussed.
\thispagestyle{empty}
\newpage
\addtocounter{page}{-1}

Up to now, theoretical attention has mostly been focused on models in
which susy-breaking gives comparable masses to the squarks and
gauginos.  However in some attractive types of models it happens
instead that gauginos are massless at tree-level and thus only get
masses on account of their interaction with other particles which are
massive. For instance if supersymmetry is broken in a hidden sector
and there are no gauge-singlets, dimension-3 operators are suppressed
by a factor $\frac{\mtilde}{M_{pl}}$ and thus are
negligible\footnote{See \cite{bkn} for a discussion of this point.}.
This scenario is particularly appealing from the standpoint of
phenomenology since it produces a low-energy theory which has far
fewer parameters than the usual MSSM and furthermore is subject to
verification or falsification in the next few years.  In this talk I
will summarize the results of work done in collaboration with Antonio
Masiero\cite{f:96} to compute the masses of the gluino and lightest
neutralino and chargino.  With this as motivation, I then survey the
results obtained in ref. \cite{f:95} concerning the hadron physics and
experimental situation of this scenario.  An experimental approach for
settling the question is discussed in ref. \cite{f:95}.   A few new
points are made here but mostly it is a review of refs.
\cite{f:96,f:95}, which should be consulted for most details.

In the absence of a tree-level susy- and R-symmetry breaking mass
term, the gluino gets its mass from top-stop loops and,
if all dimension-3 susy-breaking operators are absent, it is
completely determined just in terms of the stop and top masses and
$\alpha_{QCD}$.  The chargino and neutralino sectors in addition get
contributions from loops containing a higgsino or ew gaugino and a
Higgs or ew gauge boson.  Thus they depend in addition on the
characteristic masses of the Higgses and the parameter $\mu$
which governs the (supersymmetry-respecting) coupling between the two
Higgs doublets.  Given $m_t$, $\mu$, and $\mtilde$, one could
determine the Higgs masses for any particular scheme of ew symmetry
breaking.  For instance in the particularly attractive scenario in
which the Higgs potential develops a negative mass-squared term on
account of radiative corrections\cite{a-gpw,ir}, there is only one
free parameter in addition to $m_t$, $\mu$, and $\mtilde$, namely $B$.
However in any case the mass scale of the heavy Higgses should be of
the same order as $\mtilde$.  Thus identifying these scales, one can
estimate the gluino, chargino and neutralino masses in terms of the
two unknowns $\mu$ and $\mtilde$.

To summarize ref. \cite{f:96}:
\begin{enumerate}
\item  There are two distinct allowed regions for $\mu$ and $\mtilde$ --
namely $\mu \lsi 100$, or $\mu \gsi$ several TeV with $\mtilde \gsi 8
$ TeV -- which are consistent with the lightest chargino and squark
being more massive than the LEP lower bound of 45 GeV\footnote{It is
not appropriate to use the CDF bound\cite{cdf:gluinolim2} without
further analysis because that relies on model-dependent assumptions
about the gluino.}.  If a chargino is not discovered at LEPII with mass
less than $m_W$, the low $\mu$ region will be ruled out.
\item  For the low $\mu$ region, the mass of the gluino is $\lsi 300$
MeV. If for some reason the tree-level gaugino masses vanish but not
the dimension-3 susy-breaking squark-squark-Higgs coupling, $A$, then
the gluino mass can be higher.  E.g., if $A=1$ the gluino mass is of
order 1-2 GeV.
\item  In the large $\mu$ region the gluino mass vanishes if $A = 0$.
However if $A=1$ and $\mtilde = 10$ TeV, $m_{\gluino} \sim 40 $GeV, for
example.
\item  In the low $\mu$ region the lightest neutralino has a mass
ranging up to 400-700 MeV if $A=0$ and up to 700-1000 MeV for $A=1$.
In the low $\mu$ region the lightest neutralino is very nearly a photino.
\item  In the large $\mu$ region the lightest neutralino has a mass of
at least 10 GeV, independent of $A$, and it is almost a pure bino.
\item  Generically, the lightest neutralino is heavier than the gluino
if all dimension-3 soft-susy-breaking operators are absent from the
low energy theory, i.e., if $A=0$.
\end{enumerate}

Turning now to the hadron phenomenology of the above situation, I am
summarizing results of ref. \cite{f:95}.  If gluinos have properties
such that they decay to a photino before hadronization degrades their
energy, they can be ruled out for a very large range of masses: up to
126 GeV in a simple SUSY scenario\cite{cdf:gluinolim2}. This largely
rules out gluinos having a lifetime less than $\sim 2 \times 10^{-11}
\left( \frac{m_{\gluino}}{1{\rm GeV}}\right) $ sec.  However in the
scenario at hand the lifetime of the gluino is typically longer than
this because of phase space suppression.

An inevitable consequence of the existance of a long-lived gluino is
the existance of neutral hadrons containing them.  Generically,
hadrons containing a single gluino are called $R$-hadrons\cite{f:23}.
The lightest of these would be the neutral, flavor singlet $g \gluino$
``glueballino'', called $R^0$.  There would also be $R$-mesons,
$\bar{q}q \gluino$, and $R$-baryons,$qqq \gluino$, with the $\bar{q}q$
or $qqq$ in a color octet.  Unlike ordinary baryons which are unable
on account of fermi statistics to be in a flavor singlet state, there
is a neutral flavor-singlet $R$-baryon, $uds \tilde{g}$, called $S^0$
below.  It should be particularly strongly bound by QCD hyperfine
interactions, and probably is the lightest of the
$R$-baryons\cite{f:51,f:52}, even lighter than the $R$-nucleons.

By considering the multiplet structure of supersymmetric pure glue
QCD, and using the lattice gauge theory estimate of the mass of the
$0^{++}$ glueball in quenched approximation, the mass of the $R^0$ is
estimated in ref. \cite{f:95} to be $1440\pm375$ MeV when the gluino
is massless.  However if the gluino were massless, the spectrum would
be expected to contain an unacceptably light\cite{ev,sv}
flavor-singlet goldstone boson associated with the spontaneous
breaking of the non-anomalous linear combination of quark and gluino
chiral U(1) symmetries.  For three light flavors of quarks the
non-anomalous axial current is:
\eq
J^5_{\mu} = \frac{1}{\sqrt{26}} \left\{\bar{q}^{i,j}_L \gamma_{\mu}
q^{i,j}_L - \bar{q^c}^{i,j}_L \gamma_{\mu} {q^c}^{i,j}_L -
\bar{\lambda^a} \gamma_{\mu} \lambda^a \right\}.
\label{j5nonanom}
\en
Thus the minimum mass of the gluino can be found by requiring it to be
heavy enough to make a sufficient contribution to the mass of the
$\eta'$.  Using standard current algebra arguments\cite{f:95} this
requires
\eq
m_{\gluino} <\bar{\lambda} \lambda> \sim 11 m_s <\bar{s} s>.
\en
Since the QCD attractive force between color octets is greater
than that between triplet and antitriplet, $<\bar{\lambda} \lambda>$
is presumably larger than $<\bar{s}s>$.  Most-attractive-channel
arguments\cite{drs} suggest that the condensates depend exponentially
on the Casimirs of the condensing fermions so that since $C_8/C_3 =
9/4$, $<\bar{\lambda} \lambda>$ could be an order of magnitude or more
larger than $<\bar{s}s>$.  Thus pending lattice calculations of
$<\bar{\lambda} \lambda>$ or $m(\eta')$ as a function of gluino mass
and without gluinos, the phenomenological analysis should be general
enough to include a gluino as light as $\sim 100$ MeV or less.  In
this case the $R$-hadron properties are about the same as they would
be for a massless gluino.

The flavor singlet pseudoscalar orthogonal to the $\eta'$ which gets
its mass from the anomaly would be identified with a more massive
state.  Neglecting its quark component it is the SUSY partner of the
$R^0$ so its mass should be comparable to that of the $R^0$, i.e.,
$~1440$ MeV for a massless gluino.  Let us call this particle the
$\tilde{\eta}$.  There is evidence for an ``extra'' flavor singlet
pseudoscalar present in the meson spectrum in the 1410-1490
region\cite{PDG,mark3,dm2}, which has a large coupling to
gluons\cite{f:93}.  It is an excellent candidate for this state if it
is confirmed.

The scenario in which the $\eta'$ is a pseudogoldstone boson and a
heavier particle, the $\tilde{\eta}$, is the particle which gets its
mass from the anomaly, is attractive from the large $N_c$ point of
view.  While no-one would insist that $N_C=3$ is large enough that all
leading large-$N_c$ predictions should be valid, it has nonetheless
been astonishing to what extent the large $N_c$ limit seems to be
``precociously'' attained in hadron properties, apart from the $\eta'$
mass.  As shown by Witten\cite{witten:largeN}, in a theory in which
the anomaly only gets contributions from fermions in the fundamental
representation of the color group, the mass of the pseudoscalar which
is {\it not} a pseudogoldstone boson and gets its mass via the anomaly
must vanish as $N_c \rightarrow \infty$, while the pseudogoldstone
bosons have finite masses in this limit.  When specialized to $N_c=3$,
it leads for instance to Georgi's inequality\cite{georgi}
$\frac{m_{\eta}}{m_{\eta'}} < 0.540$, in disagreement with the
experimental value 0.572.  However in the present scenario the large
$N_c$ mass hierarchy is not violated by the pseudoscalars because
neither the $\eta'$ which is a pseudogoldstone boson, nor the
$\tilde{\eta}$ which gets its mass from the anomaly, are predicted to
have vanishing masses in the $N_c \rightarrow \infty$ limit, since
gluino loops have the same large $N_c$ behavior as gluon loops.

If the $\eta'$ were indeed the pseudogoldstone boson related to the
current in eqn (\ref{j5nonanom}), its quark content would be reduced
by a factor of $\frac{18}{26} \approx 0.7$ in comparison to the usual
picture.  Interestingly, this seems not to be ruled out by existing
constraints.  Sound predictions for the $\eta'$, avoiding model
dependent assumptions such as the relation between $F_1$ and $F_8$,
are for ratios of branching fractions to final states which couple to
the quark component\cite{chanowitz:etaprime}.  These ratios are
insensitive to the presence of a gluino or gluonic component.
Absolute predictions are highly sensitive to theoretically
incalculable hadronic effects, due to the very restricted phase space
for the $\eta'$ to decay through strong interactions.  This means that
rates which could potentially determine whether the $\eta'$ has a
$30\%$ gluino component, in practice cannot be predicted reliably
enough to be useful.\footnote{A possible way to discriminate is to
study the production of the various pseudoscalars in $J/\Psi$ decay.}
However it would be a worthwhile project to reanalyze the experimental
and theoretical constraints on the $\eta$,$\eta'$,$\tilde{\eta}$
system to see if it can be described as well in the light-gluino
interpretation as in the usual one.\footnote{Actually, the
$\tilde{\eta}$ is completely mysterious in the usual picture, so the
real question is whether in the light-gluino picture there is any
problem with finding a consistent choice of mixing angles, as was done
in ref. \cite{gk} for the conventional description.}

Let us turn now to experimental constraints on this scenario.  A
gluino in the mass range $\sim 1.5 - 3.5$ GeV is excluded, whatever
its lifetime, from the absence of a peak in the photon energy spectrum
in radiative Upsilon decay.  This is because two gluinos with mass in
that range would form a pseudoscalar bound state, the
$\eta_{\tilde{g}}$, whose branching fraction in $\Upsilon \rightarrow
\gamma \eta_{\tilde{g}}$ can be reliably computed using perturbative
QCD and is predicted\cite{keung_khare,kuhn_ono,goldman_haber} to be
greater than the experimental upper
bound\cite{tutsmunich,cusb}\footnote{The range excluded by the CUSB
experiment is incorrectly claimed to extend to lower gluino masses, by
using the pQCD results of refs.
\cite{keung_khare,kuhn_ono,goldman_haber} out of their range of
validity.  A detailed analysis of the actual excluded range in given
in ref. \cite{f:93}.  The lower limit for validity of a pQCD,
non-relativistic potential model description of an $\etagluino$ was
taken to be $\sim 3$ GeV, mainly by analogy with the success of the
same description of charmonium.  However since the effective value of
the coupling is so much stronger due to the larger color charge of the
gluino in comparison to a quark, even a 3 GeV $\etagluino$ may not be
in the perturbative regime, in which case the range of validity of the
CUSB procedure may not be even this large.  Note that any gluino whose
lifetime is longer than the strong interaction disintegration time of
the $\etagluino$, i.e., $\tau \gsi \sim 10^{-22}$ sec, will produce
the requisite bump in the photon energy spectrum, and thus be
excluded by CUSB.}.

{}From the CUSB experiment, we infer that the $\eta_{\gluino}$ does not
lie in the 3-7 GeV range, so that the gluino would not be in the $\sim
1.5-3.5$ GeV range.  In order to compare to limits from other
experiments searching for $R^0$'s, we shall convert this limit to an
effective gluino mass using the relation
\begin{equation}
m(R^0) = 0.72 (1 +
e^{-\frac{m_{\gluino}}{2}}) + m_{\gluino} (1 - e^{-m_{\gluino}}),
\label{mRmg}
\end{equation}
with all
masses in GeV. This is actually just a convention for making the
figure, but is physically reasonable in that it yields the
$m_{\gluino}=0$ result of the previous section and in analogy with
mesons made of one light and one heavy quark associates an additive
confinement energy of about half the mass of a light-quark-meson
(here, of the $0^{++}$ glueball whose mass is $\sim 1.44 $ GeV) to the
light constituent (here, the gluon) of a light-heavy composite.

Other experimental constraints are reviewed in ref. \cite{f:95}.  To
summarize that discussion: Gluinos in the mass range $\sim 1.5-3.5$
GeV are absolutely excluded (CUSB).  Lighter gluinos are allowed, as
long as the $R^0$ lifetime is not in the range $2 \times 10^{-6}-
10^{-8}$ sec if the $R^0$ mass is greater than 1.5 GeV (Bernstein et
al), or the range $> 10^{-7}$ sec if its mass is greater than 2 GeV
(Gustafson et al).  Gluinos with mass around 4 GeV or above, must have
a lifetime longer than about $\sim 2 \times 10^{-11} \left(
\frac{m_{\gluino}}{1 {\rm GeV}}\right) $ sec (UA1,CDF), with the
ranges $> 10^{-7}$ sec (Gustafson), $2 \times 10^{-6}-10^{-8}$ sec
(Bernstein et al) and $\sim 10^{-10}$ sec (ARGUS) ruled out for masses
in the vicinity of 4-5 GeV.  The figure is an attempt to summarize
these results, combining experiments which report results directly in
terms of $m(R^0)$ with those characterized by limits on $m_{\gluino}$
by use of eqn.  (\ref{mRmg}).  Given the primitive nature of
eqn. (\ref{mRmg}) and the $\pm 375$ MeV uncertainty on the $R^0$ mass
when the gluino is massless, as well as the very
rough methods used to extract the ranges of mass and lifetime
sensitivity for the various experiments, a $\gsi 20 \%$ uncertainty
should be attached to all the boundaries shown in this figure.

Some discussion of $R^0$ lifetimes which can be expected for given
gluino and lightest neutralino masses can found in ref. \cite{f:95}.
The general conclusion is that existing experimental constraints are
insufficient to exclude the particularly interesting range $1.1 <
m(R^0) < 2.2$ GeV.  Unfortunately, much more theoretical work is
needed to obtain a reliable estimate of the lifetime of the $R^0$ when
the gluino mass is $\lsi 300$ MeV but the lightest neutralino mass is
$\sim 1$ GeV (possibly the most interesting range).  Experiments which
can find or exclude such a possibility must be sensitive to long-lived
$R^0$'s, and therefore should look for its reinteraction rather than
its decay in order to be insenstive to its lifetime.  Some
experimental possibilities are discussed in ref. \cite{f:95}.
Experiments to definitively rule out or discover them are possible but
very challenging.

In the $A=0$ large $\mu$ scenario the lightest neutralino has a mass
$\gsi 10$ GeV but the gluino mass vanishes.  In this case the $R^0$
and $S^0$ would be absolutely stable unless the gravitino mass is low
enough that it provides a decay channel.  Absolute stability is a real
possibility for the $S^0$ even in the low $\mu$ region, since the mass
difference between it and the lightest neutralino must be greater than
938.8 MeV for it to decay.  If either the $R^0$ or $S^0$ bind to
nuclei, then their absolute stability could be ruled out
experimentally by the sensitive searches for exotic isotopes, at least
for some mass regions\cite{muller}.  However one would expect a
repulsive, not attractive, interaction between the flavor-singlet
$S^0$ or $R^0$ and a nucleus\cite{f:95}.  Anomalous signals in
extensive air showers and underground muons seemingly coming from
Cygnus X-3 are consistent with the intermediate particle being a
neutron, except that the neutron decays too quickly to make the long
trip\footnote{See, e.g., ref. \cite{bei} for a summary.}.  Long-lived
$R^0$'s were investigated\cite{bi}, but discarded\cite{ov} on account
of the mistaken belief that they would imply a long lived charged
$R$-proton which is ruled out by, e.g., ref. \cite{muller}.  If the
present quiet of Cygnus X-3 is only a cyclical phenomenon and such
events are observed again in the future, an $R^0$ or $S^0$
interpretation should be seriously considered.

In the usual scenario with $A \sim 1$ and tree level gaugino masses,
cosmological considerations rule out the existance of stable
neutralinos having mass less than a few GeV\footnote{See, e.g.,
ref. \cite{drees,roszkowski}.}, since in that case they would
overclose the universe.  The question needs to be revisited making
assumptions appropriate to the present scenario.  One important
difference from the usual situation in which gluinos are assumed to be
much heavier than the lightest neutralino is that in addition to the
usual reactions $\chi \chi \rightarrow f \bar{f}$ considered when
computing the annihilation of neutralinos, in this scenario one should
also include $\gluino \chi \rightarrow q \bar{q}$.  Not only is the
cross section parametrically larger by a factor $\sim
\frac{\alpha_3}{\alpha_2}$ but the reaction can go through an s-wave
because the gluino and neutralino are not identical fermions.  When
only $\chi \chi$ annihilation is relevant, the following argument
applies: a pair of self-conjugate fermions such as $\chi \chi$ or $\chi
\gluino$ state has $CP = (-1)^{L+1}$.  Meanwhile, a $\chi \chi$ state by
fermi statistics must have $(-1)^{S+1}(-1)^L = -1$ so that $L+S$ must be
even.  Hence the s-wave $\chi \chi$ state must have $S=0$ and therefore
has $J=0$. However the $J=0$ states of $f \bar{f}$ have either $L=S=0$
or $L=S=1$ and are therefore $CP$ even, since for an identical $f
\bar{f}$ pair $CP = (-1)^{L+S}$.  Thus $CP$ conservation does not allow
a $\chi \chi$ pair to annihilate to an $f \bar{f}$ pair through the
s-wave. This means that the leading term in velocity expansion of the
annihilation cross section $\sim v^2$\cite{goldberg} leading to a
significant reduction in the annihilation rate.  However for a $\chi
\gluino$ initial state one does not have the requirement from fermi
statistics that $L+S$ be even, so that they can be in an $L=0,S=1$
state whose $J=1$ is compatible with the $f \bar{f}$ being $CP$ odd.
Thus the reaction $\chi \gluino \rightarrow q \bar{q}$ is not
suppressed as $v \rightarrow 0$ and one can expect that the
annihilation of neutralinos will be much more efficient in this
scenario, reducing the cosmological limit on neutralino mass
substantially.

If the overclosure bound on the mass of a stable neutralino still
turns out to be significantly greater than 1 GeV, it could be
satisfied in the large $\mu$ region.  However this would not imply
that the large $\mu$ solution is favored, because as long as
the lightest neutralino is heavier than the $R^0$ it will decay to
$R^0$'s so their relic density will be the issue.  Another amusing
possibility which arises naturally in the low-$\mu$ region is that the
$\chi^0_1$ is heavier than the gluino, but not heavier than the $R^0$,
so that above the QCD phase transition $\chi^0_1$'s decay into
gluinos, but after the QCD phase transition the relic
$R^0$'s\footnote{Gluinos annihilate efficiently due to their strong
interactions, like the antiquarks which were present in the early
universe before they annihilated with quarks.} decay into
$\chi^0_1$'s, leaving a relic density which could account for dark
matter with photino masses characteristic of the low-$\mu$
region, i.e., $\lsi 1$ GeV.  Detailed analysis is required to assess
the quantitative viability of this suggestion. Thermal effects above
the QCD transition would contribute to an effective mass for the
gluino, but not significantly for the lightest neutralino, so that
they must be included in the calculation to draw reliable conclusions.

In summary, the possiblity of gluinos having masses less than a few
hundred Mev with the lightest neutralino mass being $\lsi 1$ GeV
emerges naturally in low energy theories without dimension-3
susy-breaking operators.  We have seen that such a scenario is not
ruled out by laboratory experiments, except for limited regions of
parameter space.  It probably is capable of accounting for the dark
matter of the universe, for values of the mass scale $\mu$
which would imply a chargino light enough to be found at LEPII.
\newpage


\begin{figure}
\epsfxsize=\hsize
\epsffile{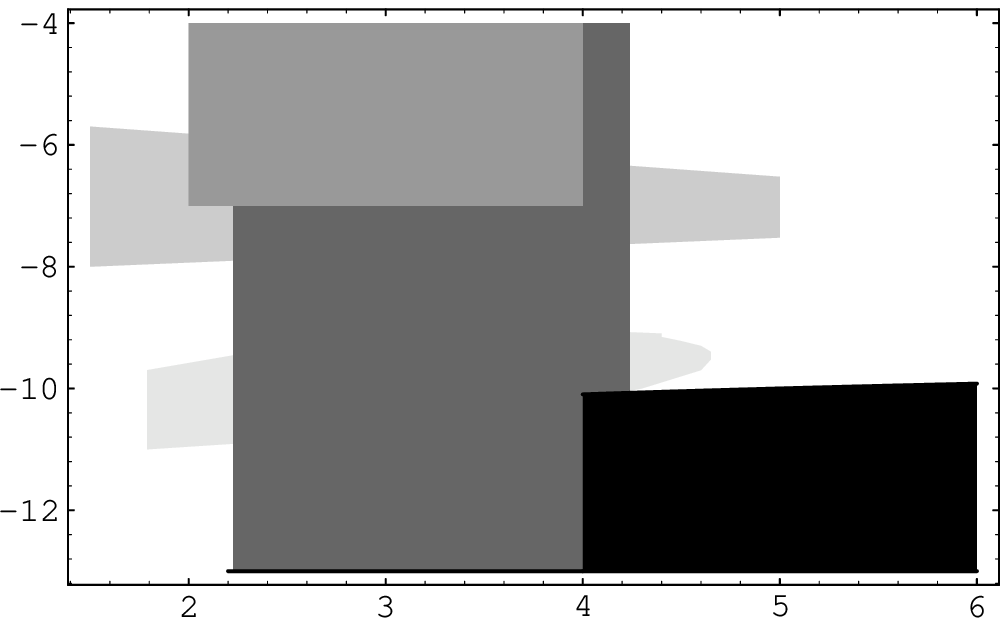}
\caption{Experimentally excluded regions of $m(R^0)$ and
$\tau_{\gluino}$.  Horizontal axis is $m(R^0)$ in GeV beginning at 1.5
GeV;  vertical axis is $Log_{10}$ of the lifetime in sec.  A massless
gluino would lead to $m(R^0) \sim 1.4 \pm .4$ GeV.  ARGUS and
Bernstein et al give the lightest and next-to-lightest regions (lower
and upper elongated shapes), respectively.  CUSB gives the
next-to-darkest block; its excluded region extends over all lifetimes.
Gustafson et al gives the smaller (mid-darkness) block in the upper
portion of the figure; it extends to infinite lifetime.  UA1 gives the
darkest block in the lower right corner; it extends to higher masses
and shorter lifetimes not shown on the figure.}
\label{ltgl}
\end{figure}

\end{document}